\documentclass[showpacs,preprintnumbers,twocolumn,
amsmath,amssymb,groupedaddress,superscriptaddress]{revtex4}

\usepackage{graphicx}
\usepackage{dcolumn}
\usepackage{bm}

\begin{document}

\title{Nuclear symmetry energy components and their
ratio: A new approach within the coherent density fluctuation
model}

\author{M. K. Gaidarov}
\affiliation{Institute for Nuclear Research and Nuclear Energy,
Bulgarian Academy of Sciences, Sofia 1784, Bulgaria}

\author{E. Moya de Guerra}
\affiliation{Grupo de F\'{i}sica Nuclear, Departamento de Estructura de la Materia (EMFTEL),\\
Facultad de Ciencias F\'{i}sicas, Universidad Complutense de
Madrid, E-28040 Madrid, Spain}

\author{A. N. Antonov}
\affiliation{Institute for Nuclear Research and Nuclear Energy,
Bulgarian Academy of Sciences, Sofia 1784, Bulgaria}

\author{I. C. Danchev}
\affiliation{Department of Physical and Mathematical Sciences,
School of Arts and Sciences, University of Mount Olive, 652 R.B.
Butler Dr., Mount Olive, NC 28365, USA}

\author{P. Sarriguren}
\affiliation{Instituto de Estructura de la Materia, IEM-CSIC,
Serrano 123, E-28006 Madrid, Spain}

\author{D. N. Kadrev}
\affiliation{Institute for Nuclear Research and Nuclear Energy,
Bulgarian Academy of Sciences, Sofia 1784, Bulgaria}

\begin{abstract}
A new alternative approach to calculate the ratio of the surface
to volume components of the nuclear symmetry energy is proposed in
the framework of the coherent density fluctuation model (CDFM). A
new expression (scheme II) for the ratio is derived consistently
within the model. This expression appears in a form more direct
and physically motivated than the expression (scheme I) that was
used in our previous works within the CDFM and avoids preliminary
assumptions and mathematical ambiguities in scheme I. The
calculations are based on the Skyrme and Brueckner energy-density
functionals for nuclear matter and on nonrelativistic
Brueckner-Hartree-Fock method with realistic Bonn B and Bonn CD
nucleon-nucleon potentials. The approach is applied to isotopic
chains of Ni, Sn, and Pb nuclei using nuclear densities obtained
in self-consistent Hartree-Fock+BCS calculations with SLy4 Skyrme
effective interaction. The applicability of both schemes within
the CDFM is demonstrated by a comparison of the results with the
available empirical data and with results of other theoretical
studies of the considered quantities. Although in some instances
the results obtained for the studied ratio and the symmetry energy
components are rather close in both schemes, the new scheme II
leads to more realistic values that agree better with the
empirical data and exhibits conceptual and operational advantages.
%It is shown that they lead to realistic values of the
%studied ratio, as well as of the components of the nuclear
%symmetry energy. It is pointed out, however, that the new scheme
%II in which the general CDFM procedure is applied without
%assumptions leads to more realistic results that are in agreement
%with the empirical data.
\end{abstract}

\pacs{21.60.Jz, 21.65.Ef, 21.10.Gv}

\maketitle

\section{Introduction}
\label{sec:intro}

The symmetry energy is a crucial quantity in nuclear physics and
its astrophysical applications (see, e.g.,
Ref.~\cite{Bombaci2018}). Also, heavy-ion collisions at
intermediate energies are considered as an unique tool to explore
the nuclear equation of state (EOS) under laboratory controlled
conditions. For instance, the combined experiment at GANIL, where
the VAMOS spectrometer was coupled with the 4$\pi$ INDRA detector
to study the isotopic distributions produced in
$^{40,48}$Ca+$^{40,48}$Ca collisions at 35 MeV/nucleon, allowed
one to estimate the relative contribution of surface and volume
terms to the symmetry energy in the nuclear EOS \cite{Chbihi2015}.
The knowledge of this contribution and, especially, the relevance
of the surface term are important to explore to what extent one
can learn about the density dependence of the symmetry energy in
infinite nuclear matter (NM) from multifragmentation of finite
nuclei and from nuclear reaction dynamics.

The density dependence of the symmetry energy is fairly unknown
and there are very different predictions for the various models
(see, for instance, Ref.~\cite{Klahn2006}). It is revealed in the
relationship between the basic symmetry energy parameters and the
neutron-skin thickness in a heavy nucleus. The latter has an
explicit dependence on the slope parameter of the symmetry energy
via the density dependence of the surface tension, which has been
determined in Ref.~\cite{Horiuchi2017} within a compressible
droplet model. Measurements of nuclear structure characteristics
including masses, densities, and collective excitations have
resolved some of the basic features of the EOS of nuclear matter.
The EOS allows one to constrain the bulk and surface properties of
the nuclear energy-density functionals (EDFs) quite effectively
via the symmetry energy and related properties. The latter are
significant ingredients of the EOS and their study in both
asymmetric nuclear matter and finite nuclei are of particular
importance.

The symmetry energy of a finite nucleus is a collective feature
with volume and surface terms, but it is also related to the
nucleon-nucleon ($NN$) interaction and the energy-density
functional. For example, there is an analytical parametrization of
the link between the Skyrme (see Ref.~\cite{Wang2015}) and
Brueckner \cite{Brueckner68,Brueckner69} EDFs and the symmetry
energy (see, e.g., \cite{Wang2015,Gaidarov2011}). Additionally,
when exploring the symmetry energy properties of neutron-rich
nuclei by means of the nonrelativistic Brueckner-Hartree-Fock
(BHF) method with modern realistic Bonn B and Bonn CD potentials
\cite{Danchev2020}, we performed nuclear matter many-body
calculations and then, results for finite nuclei are obtained
within the coherent density fluctuation model (CDFM) (e.g.,
Refs.~\cite{Antonov80,AHP}). The usage of the latter (that will be
discussed below and applied in the present work) is closely
related to the general point of the proper account for the $NN$
correlations. It is known that the short-range and tensor $NN$
correlations induce a population of high-momentum components in
the many-body nuclear wave function. In Ref.~\cite{Carbone2012}
the impact of such high-momentum components on bulk observables
associated with isospin asymmetric matter has been studied. It was
shown that the kinetic part of the symmetry energy is strongly
reduced by correlations when compared to the non-interacting case.
The results for nuclear matter obtained in Ref.~\cite{Vidana2015}
have confirmed the critical role of the tensor force in the
determination of the symmetry energy and its slope parameter even
at densities not too far above the saturation density.

The coherent density fluctuation model is a natural extension of
the Fermi gas model based on the $\delta$-function limit of the
generator coordinate method \cite{AHP,Grif57} and includes
long-range correlations of collective type. The CDFM has been
applied to calculate various quantities of nuclear structure and
reactions. Here we will mention some of them. In Ref.~\cite{Ant88}
CDFM results for the energies, density distributions and rms radii
of the ground and first monopole states in $^{4}$He, $^{16}$O and
$^{40}$Ca nuclei have been obtained. Also, results for the
incompressibility of finite nuclei have been reported in
Refs.~\cite{AHP,Ant91}. The CDFM has been employed to calculate
the scaling function in nuclei using the relativistic Fermi gas
scaling function \cite{Ant2004,Ant2005} and the result has been
applied to lepton scattering processes
\cite{Ant2004,Ant2005,Ant2006a,Ant2006b,Ant2008,Ant2007,Ant2009}.
In particular, the CDFM analyses became useful to obtain
information about the role of the nucleon momentum and density
distributions for the explanation of superscaling in
lepton-nucleus scattering \cite{Ant2005,Ant2006a}. The CDFM
scaling function has been used to predict cross sections for
several processes: inclusive electron scattering in the
quasielastic and $\Delta$ regions \cite{Ant2006b,Ant2008} and
neutrino (antineutrino) scattering both for charge-changing
\cite{Ant2008,Ant2009} and neutral-current \cite{Ant2007,Ant2009}
processes. The model was applied to study the scaling function and
its connection with the spectral function and the momentum
distribution \cite{Cab2010,Antonov2011}. The CDFM was used also to
study the role of the $NN$ correlations in elastic magnetic
electron scattering \cite{Kadrev96,Sarriguren2019,Hernandez2021}.

In our previous works
\cite{Gaidarov2011,Gaidarov2012,Gaidarov2014,Antonov2016,Antonov2018,Danchev2020,Gaidarov2020,Gaidarov2020ch}
we have demonstrated the capability of CDFM to be applied as an
alternative way to make a transition from the properties of
nuclear matter to the properties of finite nuclei investigating
the nuclear symmetry energy (NSE), the neutron pressure, and the
asymmetric compressibility in finite nuclei. While there is enough
collected information for these key EOS parameters (although the
uncertainty of their determination is still large), the volume and
surface symmetry energies have been poorly investigated till now.
This concerns mostly the surface contribution to the NSE and comes
from the fact that many nucleons are present at around the nuclear
surface. The volume and surface contributions to the NSE and their
ratio at zero temperature were calculated in
Ref.~\cite{Antonov2016} within the CDFM using two EDFs, namely,
the Brueckner and Skyrme ones. The CDFM weight function was
obtained by means of the proton and neutron densities obtained
from the self-consistent deformed Hartree-Fock (HF)+BCS method
with density-dependent Skyrme interactions. The obtained results
in the cases of Ni, Sn, and Pb isotopic chains were compared with
results of other theoretical methods and with those from
approaches which used experimental data on binding energies,
excitation energies to isobaric analog states (IAS), and
neutron-skin thicknesses. An investigation of the thermal
evolution of the NSE components and their ratio for isotopes
belonging to the same chains around the double-magic nuclei
performed in Ref.~\cite{Antonov2018} has extended our previous
analysis of these nuclei for temperatures different from zero.

In this paper, we revisit the expression for the ratio between the
volume and surface components to the NSE within the CDFM proposed
in Refs.~\cite{Danchev2020,Antonov2016} and suggest a new
alternative approach in a more direct and physically motivated way
to calculate this ratio. The main aim of the work is to avoid the
preliminary assumptions and mathematical ambiguities in our
previous scheme I. To achieve this goal, in the new scheme II, we
apply the general relation based on the Droplet Model between the
symmetry energy and its components to the building units
("fluctons") of the CDFM model, and we construct from them the
ratio between the NSE components for finite nuclei following the
standard CDFM procedure. This provides more solid physical grounds
to the new scheme that is expected to lead to more reliable
results. We also search for the dependence of the results on
several sets of nuclear potentials. In the new approach we perform
calculations for the symmetry energy components $S^{V}(A)$ and
$S^{S}(A)$ and their ratio for the same isotopes in Ni
($A=74-84$), Sn ($A=124-156$), and Pb ($A=202-214$) chains
considered before and compare the obtained results with the
previous ones (including $S^{V}(A)$, $S^{S}(A)$, and their ratio
$\kappa$) obtained by the procedure in
Refs.~\cite{Danchev2020,Antonov2016}. The applicability of our
both schemes within the CDFM is also demonstrated by a comparison
of the results with the available empirical data and with results
of other theoretical studies for the considered quantities.

In the next section, we give definitions of the EOS parameters
governing its density dependence in nuclear matter and finite
nuclei using the CDFM, as well as expressions of various
quantities of interest in the previous and new alternative
approach within the CDFM. Section~\ref{subsec:EOS} briefly
explains the derivation of the expression for the ratio of the
volume to the surface symmetry energy coefficients on the basis of
the local density approximation to the symmetry energy.
Section~\ref{subsec:CDFM} contains the previous CDFM formalism
that provides a way to calculate the mentioned quantities. The new
alternative approach aiming to calculate the NSE components and
their ratio is formulated in Sec.~\ref{subsec:newCDFM}. Our
numerical results are presented and discussed in
Sec.~\ref{sec:results}. The main conclusions of the study are
summarized in Sec.~\ref{sec:conclusions}.

\section{Theoretical scheme}
\label{sec:formalism}

\subsection{Main relationships for EOS parameters in nuclear matter and in finite
nuclei}
\label{subsec:EOS}

The Bethe-Weizs\"{a}cker semi-empirical mass formula describes
both properties of symmetric (finite) nuclear matter as well as
the essential dependence of the finite nucleus ground-state energy
on the isospin asymmetry (polarization)
\cite{vonWeiz35,Bethe71,Myers1969,Steiner2005}:
\begin{eqnarray}
E(A,Z)&=&-B+E_SA^{-1/3} + S(A)\frac{(N-Z)^2}{A^2}+ E_C \frac{Z^2}{A^{4/3}} \nonumber\\
&& +E_{dif} \frac{Z^2}{A^2}+E_{ex}\frac{Z^{4/3}}{A^{4/3}}+a\Delta
A^{-3/2}.
\label{eq:1}
\end{eqnarray}
In Eq.~(\ref{eq:1}) $B \simeq 16$ MeV is the binding energy per
particle of bulk symmetric matter at saturation. $E_{S}$, $E_{C}$,
$E_{dif}$, and $E_{ex}$ correspond to the surface energy of
symmetric matter, the Coulomb energy of a uniformly charged
sphere, the diffuseness correction, and the exchange correction to
the Coulomb energy, respectively. The last term gives the pairing
correction, $\Delta$ is a constant and $a=+1$ for odd-odd nuclei,
0 for odd-even, and -1 for even-even nuclei. $S(A)$ is the
symmetry energy expressed by the volume $S^{V}(A)$ and modified
surface component $S^{S}(A)$ in the droplet model (see
Ref.~\cite{Steiner2005}, where it is defined as $S_{s}^{*}$):
\begin{equation}
S(A)=\frac{S^V(A)}{1 + \displaystyle
\frac{S^S(A)}{S^V(A)}A^{-1/3}}= \frac{S^V(A)}{1+ \displaystyle
q(A)A^{-1/3}},
\label{eq:2}
\end{equation}
where
\begin{equation}
q(A) \equiv \frac{S^S(A)}{S^V(A)}.
\label{eq:3}
\end{equation}
We note that in the present work we use Eq.~(\ref{eq:2}) as a
basic relation between the symmetry energy $S(A)$ and its volume
$S^{V}(A)$ and surface $S^{S}(A)$ components. The reason to use
Eq.~(\ref{eq:2}) in contrast to the relation in another approach
used in, e.g., Refs.~\cite{Dan2006,Dan2004,Dan2003,Danielewicz},
and also in our work \cite{Antonov2016}, was discussed in detail
in our previous work \cite{Danchev2020}. It is motivated by the
necessity to have a correct behavior of the denominator in
Eq.~(\ref{eq:2}) in the infinite nuclear matter limit. More
precisely, in the limit $A \rightarrow \infty$ the ratio in
Eq.~(\ref{eq:2}) $S^S/S^V \rightarrow 0$, so that
$[S^{S}/S^{V}]A^{-1/3} \rightarrow 0$ and the symmetry energy in
Eq.~(\ref{eq:2}) has the correct limit $S \rightarrow S^V$.
Contrary to this, in the approach of
Refs.~\cite{Dan2006,Dan2004,Dan2003,Danielewicz} in the limit $A
\rightarrow \infty$ the term $[S^{V}(A)/S^{S}(A)]A^{-1/3}$ is not
well determined. The use of the latter approach needs a condition
to be imposed, namely the surface coefficient $S^S(A)$ to go to
zero more slowly than $A^{-1/3}$ as $A \rightarrow \infty$. This
is the reason to use in our work Eq.~(\ref{eq:2}) instead of the
relation in the approach in e.g.,
Refs.~\cite{Dan2006,Dan2004,Dan2003,Danielewicz}.

At very large $A$ we may write the symmetry energy in the known
form (see Ref.~\cite{Bethe71}):
\begin{equation}
S(A) \simeq S^V(A)-\frac{S^S(A)}{A^{1/3}},
\label{eq:4}
\end{equation}
which follows from Eq.~(\ref{eq:2}) for large $A$.

The relations of $S^{V}(A)$ and $S^{S}(A)$ with $S(A)$ in terms of
$q(A)$ can be found from Eqs.~(\ref{eq:2}) and (\ref{eq:3}):
\begin{equation}
S^{V}(A)=S(A)\left[1+\frac{q(A)}{A^{1/3}}\right],
\label{eq:5}
\end{equation}
\begin{equation}
S^{S}(A)=q(A) S(A) \left[1+\frac{q(A)}{A^{1/3}}\right].
\label{eq:6}
\end{equation}

The following expression for the ratio of the volume to the
surface symmetry energy coefficients was given by Danielewicz
\cite{Dan2006} (see also Ref.~\cite{Diep2007}):
\begin {equation}
\kappa (A)=\frac{S^V(A)}{S^S(A)}=\frac{3}{r_{0}}\int dr
\frac{\rho(r)}{\rho_{0}} \left
\{\frac{S^{NM}(\rho_{0})}{S^{NM}[\rho(r)]}-1\right \},
\label{eq:7}
\end{equation}
where $S^{NM}[\rho(r)]$ is the nuclear matter symmetry energy,
$\rho(r)$ is the half-infinite nuclear matter density, $\rho_{0}$
is the nuclear matter equilibrium density, and $r_{0}$ is the
radius of the nuclear volume per nucleon. The latter two
quantities are related by
\begin {equation}
\frac{4\pi r_{0}^{3}}{3}=\frac{1}{\rho_{0}}.
\label{eq:8}
\end{equation}

Here we give for completeness the following general expression for
the nuclear matter symmetry energy used in Eq.~(\ref{eq:7}):
\begin{eqnarray}
S^{NM}(\rho)&=&\frac{1}{2}\left.
\frac{\partial^{2}E(\rho,\delta)}{\partial\delta^{2}} \right
|_{\delta=0} = a_{4}+\frac{p^{NM}_{0}}{\rho_{0}^{2}}(\rho-\rho_0)\nonumber \\
&&+ \frac{\Delta K^{NM}}{18\rho_{0}^{2}}(\rho-\rho_{0})^{2}+
\cdots ,
\label{eq:9}
\end{eqnarray}
where $E(\rho,\delta)$ is the energy per particle for nuclear
matter that depends on the density and the isospin asymmetry
$\delta=(\rho_{n}-\rho_{p})/\rho$ with the baryon density
$\rho=\rho_{n}+\rho_{p}$, $\rho_{n}$ and $\rho_{p}$ being the
neutron and proton densities. The parameter $a_{4}$ is the
symmetry energy at equilibrium $[a_{4}=S^{NM}(\rho_{0})]$, while
the pressure $p_{0}^{NM}$ and the curvature $\Delta K^{NM}$ have
the corresponding forms:
\begin{equation}
p_{0}^{NM}=\rho_{0}^{2}\left.
\frac{\partial{S^{NM}}}{\partial{\rho}} \right |_{\rho=\rho_{0}},
\label{eq:10}
\end{equation}
\begin{equation}
\Delta K^{NM}=9\rho_{0}^{2}\left.
\frac{\partial^{2}S^{NM}}{\partial\rho^{2}} \right
|_{\rho=\rho_{0}}.
\label{eq:11}
\end{equation}

In the next two subsections we present our relationships for the
ratio of $S^{V}(A)$ and $S^{S}(A)$ obtained in the approaches
considered within the framework of the coherent density
fluctuation model \cite{Antonov80,AHP,Antonov2016,Danchev2020}.
Our results for the mentioned quantities are given in
Sec.~\ref{sec:results}.

\subsection{EOS parameters of finite nuclei in the CDFM}
\label{subsec:CDFM}

In the present work we calculate the EOS parameters in finite
nuclei, such as the nuclear symmetry energy and its surface and
volume components using the CDFM. As mentioned in the
Introduction, the model is based on the $\delta$-function limit of
the generator coordinate method \cite{AHP,Grif57}, it is a natural
extension of the Fermi-gas model and includes $NN$ correlations of
collective type. An important feature of the CDFM is that it
allows us to make the transition from nuclear matter quantities to
the corresponding ones in finite nuclei. In the CDFM the one-body
density matrix $\rho(\mathbf{r},\mathbf{r}^{\prime})$ is a
coherent superposition of the one-body density matrices
$\rho^{NM}_{x}({\bf r},{\bf r^{\prime}})$ for spherical ``pieces''
of nuclear matter (``fluctons'') with radius $x$ and density
$\rho_{x}({\bf r})=\rho_{0}(x)\Theta(x-|{\bf r}|)$, where
\begin{equation}
\rho_{0}(x)=\frac{3A}{4\pi x^{3}},
\label{eq:11a}
\end{equation}
in which all A nucleons are homogeneously distributed:
\begin{eqnarray}
\rho^{NM}_{x}({\bf r},{\bf r^{\prime}})&=&3\rho_{0}(x)
\frac{j_{1}(k_{F}(x)|{\bf r}-{\bf r^{\prime}}|)}{(k_{F}(x)|{\bf
r}-{\bf r^{\prime}}|)}\nonumber \\  & \times & \Theta \left
(x-\frac{|{\bf r}+{\bf r^{\prime}}|}{2}\right ).
\label{eq:12}
\end{eqnarray}
It has the form:
\begin{equation}
\rho({\bf r},{\bf r^{\prime}})=\int_{0}^{\infty}dx |F(x)|^{2}
\rho^{NM}_{x}({\bf r},{\bf r^{\prime}}).
\label{eq:13}
\end{equation}
In Eq.~(\ref{eq:12}) $j_{1}$ is the first-order spherical Bessel
function and
\begin{equation}
k_{F}(x)=\left(\frac{3\pi^{2}}{2}\rho_{0}(x)\right )^{1/3}\equiv
\frac{\beta}{x}
\label{eq:14}
\end{equation}
with
\begin{equation}
\beta=\left(\frac{9\pi A}{8}\right )^{1/3}\simeq 1.52A^{1/3}
\label{eq:15}
\end{equation}
is the Fermi momentum of the nucleons in the flucton. The nucleon
density distribution in the CDFM has the form:
\begin{equation}
\rho({\bf r})=\int_{0}^{\infty}dx
|F(x)|^{2}\rho_{0}(x)\Theta(x-|{\bf r}|).
\label{eq:16}
\end{equation}
It can be seen from Eq.~(\ref{eq:16}) that in the case of
monotonically decreasing local density ($d\rho/dr\leq 0$) the
weight function $|F(x)|^{2}$ can be obtained from a known density
(obtained theoretically or experimentally):
\begin{equation}
|F(x)|^{2}=-\frac{1}{\rho_{0}(x)} \left. \frac{d\rho(r)}{dr}\right
|_{r=x} .
\label{eq:17}
\end{equation}
It has been shown in our previous works
\cite{Antonov2016,Gaidarov2011,Gaidarov2012} that the following
expression for the nuclear symmetry energy in finite nuclei $S(A)$
can be obtained within the CDFM on the base of the infinite matter
one $S^{NM}(\rho)$ (at temperature $T=0$ MeV)  by weighting it
with $|F(x)|^{2}$:
\begin{equation}
S(A)=\int_{0}^{\infty}dx|F(x)|^{2}S^{NM}[\rho(x)].
\label{eq:18}
\end{equation}
Here we would like to note that when our procedure is applied to
quantities of (infinite) nuclear matter, the self-consistency
requires the weight function to reduce to Dirac $\delta$-function.
For instance, when the self-consistency is applied to the density
$\rho(|{\bf r}|)$ and the symmetry energy $S^{NM}[\rho(|{\bf
r}|)]$ in nuclear matter it leads from Eqs.~(\ref{eq:16}) and
(\ref{eq:18}) to the identities:
\begin{eqnarray}
\rho^{NM}(|{\bf r}|,x)&=&\int_{0}^{\infty}dx' \delta(x'-x)\rho_{0}(x')\Theta(x'-|{\bf r}|)\nonumber\\
&=& \rho_{0}(x)\Theta(x-|{\bf r}|),
\label{eq:19}
\end{eqnarray}
\begin{eqnarray}
S^{NM}[\rho^{NM}(|{\bf r}|,x)]&=&\int_{0}^{\infty}dx' \delta(x'-x)S^{NM}[\rho^{NM}(|{\bf r}|,x')]\nonumber\\
&=&S^{NM}[\rho_{0}(x)\Theta(x-|{\bf r}|)].
\label{eq:20}
\end{eqnarray}

In our already mentioned works (including Ref.~\cite{Danchev2020})
we applied the CDFM in the framework of the self-consistent
Skyrme-Hartree-Fock plus BCS method to calculate the volume and
surface components of the symmetry energy and their ratio in the
Ni, Sn, and Pb isotopic chains. In our first scheme to calculate
the ratio $\kappa(A)$ we started from the expression of
Eq.~(\ref{eq:7}) (see, e.g., \cite{Dan2006,Diep2007}) making in it
a preliminary assumption replacing the density $\rho(r)$ for the
half-infinite nuclear matter in the integrand by the density
distribution of a finite nucleus, namely, by the expression in the
CDFM [Eq.~(\ref{eq:16})]. Following the procedure whose details
are given in our work \cite{Danchev2020} and using
Eqs.~(\ref{eq:19}) and (\ref{eq:20}), we obtain the formula for
$\kappa(A)$ in the form:
\begin{eqnarray}
\kappa(A) &=& \frac{3}{r_{0}\rho_{0}}\int_{0}^{\infty}dx
|{\cal F}(x)|^{2} \rho_{0}(x) \nonumber \\
&&\times\int_{0}^{x}dr\left
\{\frac{S^{NM}(\rho_{0})}{S^{NM}[\rho_0(x)]}-1\right \}
\label{eq:21}
\end{eqnarray}
that leads finally to
\begin{equation}
\kappa(A)=\frac{3}{r_{0}\rho_{0}}\int_{0}^{\infty}dx |{\cal
F}(x)|^{2} x \rho_{0}(x)\left
\{\frac{S^{NM}(\rho_{0})}{S^{NM}[\rho_0(x)]}-1\right \} .
\label{eq:22}
\end{equation}
The right-hand side of Eq.~(\ref{eq:22}) is an one-dimensional
integral over $x$, the latter being the radius of the ``flucton''
that is perpendicular to the nuclear surface. We refer to the
expression in Eq.~(\ref{eq:22}) as scheme I, because this was the
first equation that we used for the numerical calculations of the
results presented in \cite{Danchev2020, Antonov2016}. We note that
a careful analysis of the integration interval in
Eq.~(\ref{eq:22}) required in order to avoid possible
singularities in the integrand in some $x$ ranges was carried out
in our previous works.

\subsection{An alternative approach in the CDFM to calculate the ratio of surface to volume
components of the nuclear symmetry energy}
\label{subsec:newCDFM}

As mentioned in the Introduction, the main aim in the present work
is to provide a new scheme to calculate the ratio $q(A)$ as
defined in Eq.~(\ref{eq:3}). Here we would like to underline the
main differences in the construction of scheme II in comparison
with the previous scheme I: i) we do not use the method in
Refs.~\cite{Dan2006,Diep2007}, and ii) we avoid the above
mentioned assumption in subsection \ref{subsec:CDFM}, namely the
replacement of the density $\rho(r)$ for the half-infinite nuclear
matter by the density distribution of a finite nucleus. A third
and important reason to choose a new scheme is that the integrand
in Eq.~(\ref{eq:22}) for $\kappa$ in scheme I presents
singularities for some of the potentials (e.g., for the Brueckner
one). Thus, the results for $\kappa$ become extremely sensitive to
the choice of the integration interval, mainly to the value of the
lower limit of integration in Eq.~(\ref{eq:22}). In the new scheme
II we start from the general relationship [Eq.~(\ref{eq:2})]
between the NSE $S$ and its components $S^{V}$ and $S^{S}$. The
procedure of the derivation of $q(A)$ for finite nuclei is as
follows: i) we determine the ratio $q(x)=S^{S}(x)/S^{V}(x)$ for
the ``fluctons'' of the CDFM from the basic Eqs.~(\ref{eq:2}) and
(\ref{eq:4}), and ii) we construct $q(A)$ within the CDFM rules
weighting $q(x)$ by the weight function $|F(x)|^{2}$. First, to
construct $q(x)=S^{S}(x)/S^{V}(x)$ in the $x$-flucton we recall
that the $x$-flucton is a sphere of nuclear matter of radius $x$
with density $\rho_{0}(x)$. This implies that inside each flucton
we may apply Eq.~(\ref{eq:4}) in the form $S^{S}/S^{V}\simeq (1-
S/S^{V})A^{1/3}$, with $A$, the number of nucleons in the flucton,
given by $(x/r_{0})^{3}[\rho_{0}(x)/\rho_{0}]$ [see
Eqs.~(\ref{eq:8}) and (\ref{eq:11a})], and $S$ the nuclear matter
symmetry energy in the flucton $[S^{NM}(\rho_{0}(x))]$ with volume
component $S^{V}\simeq S^{NM}(\rho_{0})$. This results in the
following expression for $q(x)$:
\begin{equation}
q(x)=\frac{S^{S}(x)}{S^{V}(x)}=\frac{x}{r_{0}}\left[\frac{\rho_{0}(x)}{\rho_{0}}\right]^{1/3}
\left[1-\frac{S^{NM}[\rho_{0}(x)]}{S^{NM}(\rho_{0})}\right].
\label{eq:24}
\end{equation}
Weighting $q(x)$ by the  function $|F(x)|^{2}$ leads to the
following relationship for the ratio (\ref{eq:3}):
\begin{eqnarray}
q(A)&=&\int_{0}^{\infty}dx |F(x)|^{2}q(x)=\int_{0}^{\infty}dx
|F(x)|^{2}\nonumber \\
&\times&
\frac{x}{r_{0}}\left[\frac{\rho_{0}(x)}{\rho_{0}}\right]^{1/3}
\left[1-\frac{S^{NM}[\rho_{0}(x)]}{S^{NM}(\rho_{0})}\right].
\label{eq:25}
\end{eqnarray}
We refer to the expression in Eq.~(\ref{eq:25}) as scheme II. Here
we would like to note the following: i) the expression
Eq.~(\ref{eq:24}) for a flucton is obtained in a direct and
natural way starting from the known formula Eq.~(\ref{eq:4}) that
follows from the general relationship Eq.~(\ref{eq:2}) at large
$A$; ii) Eq.~(\ref{eq:25}) is obtained without preliminary
assumptions that were imposed to obtain Eq.~(\ref{eq:22}) in
scheme I and is free from singularities; iii) as a result of i)
and ii) the calculated quantity $1/q=S^{V}/S^{S}$ that follows
from Eq.~(\ref{eq:25}) is not equal to the previously calculated
quantity $\kappa$ following Eq.~(\ref{eq:22}). We note that both
quantities are obtained within different schemes, though both are
within the framework of the CDFM. Of course, the values of the
results for $1/q(A)$ coming from Eq.~(\ref{eq:25}) and $\kappa(A)$
[Eq.~(\ref{eq:22})] can be compared and this is done in the next
section, analyzing in this way the role of the assumptions made in
approach I and the new direct CDFM scheme II (the latter being
without extra assumptions and free from singularities) on the
studied quantities.

In the next section \ref{sec:results} we present our results for
the new ratio $q(A)$ as well as the new CDFM results for the
symmetry energy components $S^{V}(A)$ [Eq.~(\ref{eq:5})] and
$S^{S}(A)$ [Eq.~(\ref{eq:6})] in terms of $S(A)$
[Eq.~(\ref{eq:18})] and $q(A)$ [Eq.~(\ref{eq:25})], in comparison
with our previous results for the corresponding quantities in the
case of the three isotopic chains of  Ni, Sn, and Pb using Skyrme,
Bruckner, Bonn B and Bonn CD potentials. The self-consistent
Skyrme-HF plus BCS method is used in the calculations of the
nuclear densities of these nuclei and the CDFM weight function
$|F(x)|^{2}$. The results for the ratio $1/q$ are presented and
discussed in relation to the values of $\kappa$ and compared with
the available empirical data and with results of other theoretical
considerations.

\section{Results and discussion}
\label{sec:results}

As the main emphasis of the present study is to propose a new
approach to study the nuclear symmetry energy components and their
ratio, we start our analysis with the two basic quantities
entering the integrands in Eq.~(\ref{eq:25}) for the ratio $q(A)$,
namely the symmetry energy of nuclear matter $S^{NM}[\rho_{0}(x)]$
in a flucton with density $\rho_{0}(x)$ and the weight function
$|F(x)|^{2}$. Then, obtaining the symmetry energy in finite nuclei
within the CDFM from Eq.~(\ref{eq:18}) its volume and surface
components in the new approach can be calculated from
Eqs.~(\ref{eq:5}) and (\ref{eq:6}), respectively.

\begin{figure}
\centering
\includegraphics[width=80mm]{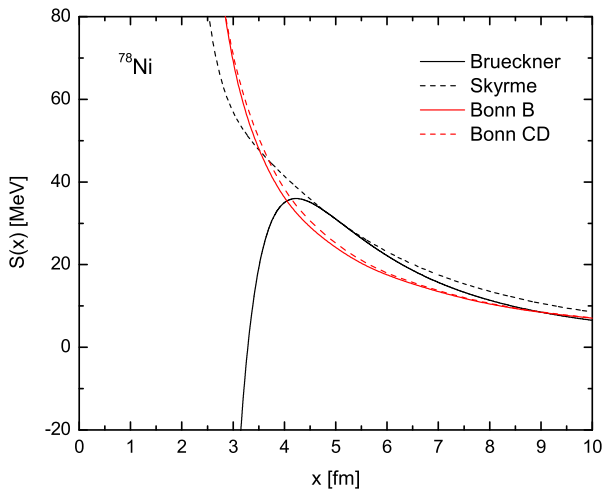}
\caption[]{(Color online) The symmetry energy $S(x)$ of the
double-magic nucleus $^{78}$Ni as a function of the flucton radius
$x$ [related to its density $\rho_{0}(x)=3A/(4\pi x^{3})$]
calculated with Brueckner EDF (black solid line), Skyrme EDF
(black dashed line) and BHF method with Bonn B (red solid line)
and Bonn CD (red dashed line) potentials from
\cite{Machleidt_ANP1989,Machleidt2001}. \label{fig1}}
\end{figure}

As an example, we show first in Fig.~\ref{fig1} the results for
the symmetry energy $S(x)$ of the double-magic nucleus $^{78}$Ni
corresponding to nonrelativistic BHF results with realistic Bonn B
and Bonn CD potentials as well as to the Brueckner and Skyrme
EDFs, for which analytical expressions for $S^{NM}(x)$ can be
found in
Refs.~\cite{Gaidarov2011,Gaidarov2012,Gaidarov2014,Antonov2016}.
As can be seen in Fig.~\ref{fig1}, the symmetry energy derived
from the energy density functional of Brueckner {\it et al.}
\cite{Brueckner68,Brueckner69} goes extremely down below $x=4$ fm
in comparison with the other three cases. The latter exhibit a
smooth behavior and their corresponding curves are close to each
other. The Brueckner symmetry energy curve behaves similarly to
them in the range $x>4$ fm. The reason for the particular behavior
of $S(x)$ in the case of Brueckner EDF lies in its parametrization
as a function of the density performed in nuclear matter
calculations. The symmetry energy versus $x$ plotted in
Fig.~\ref{fig1} corresponds to the Brueckner curve displayed in
Fig.~5 of Ref.~\cite{Danchev2020}, where the symmetry energy is
given versus the density $\rho$, as follows: the region $x\leq 4$
fm corresponds to the right "wing" after the maximum of
$S^{NM}(\rho)$ at around $\rho=0.24$ fm$^{-3}$ (see Fig.~5 of
Ref.~\cite{Danchev2020}), while the region $x>4$ fm refers to the
left "wing" before the maximum. The behavior of $S(x)$ in the case
of the Brueckner EDF shows its isospin instability. Due to this
fact, a lower cutoff is needed to compute Eqs.~(\ref{eq:18}) and
(\ref{eq:25}), but this is naturally supplied by the function
$|F(x)|^{2}$ as explained in the discussions that follow. Here we
should note that the observed differences of the symmetry energy
at $x<4$ fm in the particular case of $^{78}$Ni (see
Fig.~\ref{fig1}) provide us with a hint about the range of the
lower limit of integration in Eq.~(\ref{eq:25}) in order to get
correct physical values for the ratio of the surface to volume
components of the nuclear symmetry energy.

\begin{figure}
\centering
\includegraphics[width=80mm]{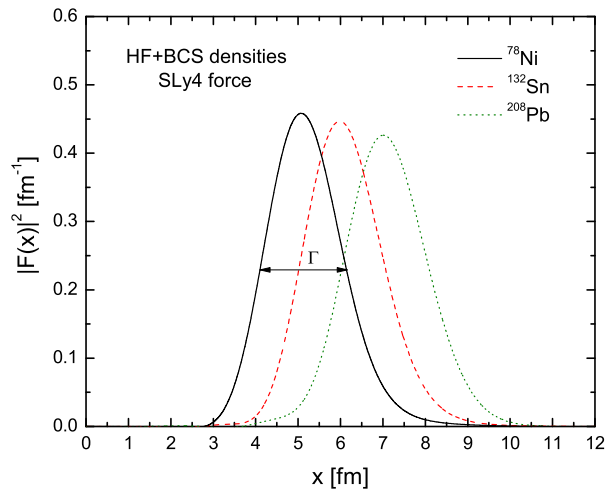}
\caption[]{(Color online) The weight functions $|F(x)|^{2}$
[Eq.~(\ref{eq:17})] of double-magic $^{78}$Ni, $^{132}$Sn, and
$^{208}$Pb nuclei calculated in the Skyrme HF+BCS method with SLy4
force.
\label{fig2}}
\end{figure}

In Fig. 2 are given the CDFM weight functions $|F(x)|^{2}$ of
double-magic $^{78}$Ni, $^{132}$Sn, and $^{208}$Pb nuclei as a
function of the flucton radius $x$. Their densities are obtained
in self-consistent HF+BCS calculations with SLy4 interaction
\cite{Chabanat98}. The function $|F(x)|^{2}$ which is used in
Eqs.~(\ref{eq:18}), (\ref{eq:22}), and (\ref{eq:25}) has the form
of a bell with a maximum around $x=R_{1/2}$ at which the value of
the density $\rho(x=R_{1/2})$ is around half of the value of the
central density equal to $\rho_{c}$
$[\rho(R_{1/2})/\rho_{c}=0.5]$. Namely, in this region around
$\rho=\rho_{c}/2$ the values of different $S^{NM}(\rho)$ play the
main role in the calculations. Therefore, to fully specify the
role of both quantities $S^{NM}[\rho_{0}(x)]$ and $|F(x)|^{2}$ in
the expression (\ref{eq:25}) for the ratio $q(A)$ and to determine
the relevant region of densities in finite nucleus calculations,
we apply a physical criterion related to the weight function
$|F(x)|^{2}$. The latter contains the nuclear structure
information through the total nuclear density. In this respect,
the width $\Gamma$ of the weight function $|F(x)|^{2}$ at its half
maximum (which is illustrated in Fig.~\ref{fig2} on the example of
$^{78}$Ni nucleus) is a good and acceptable choice.

As it is known, the central density of the nucleus has values
around $\rho_{c}\approx $ 0.10-0.16 fm$^{-3}$. Consequently, the
maximum of the weight function $|F(x)|^{2}$ is around
$\rho(R_{1/2})\approx $ 0.05-0.08 fm$^{-3}$. In the case of
$^{78}$Ni (see Fig.~6 in Ref.~\cite{Danchev2020}) the maximum of
$|F(x)|^{2}$ is at $\rho=0.05$ fm$^{-3}$ and, within its width
range, the density $\rho$ is between 0.12 fm$^{-3}$ and 0.01
fm$^{-3}$. Thus, from the combined analysis of $S^{NM}(\rho)$ and
$|F(x)|^{2}$ it turns out that the relevant values of the NM
symmetry energy are typically those in the region around $\rho
\approx $ 0.01-0.12 fm$^{-3}$ (see also the discussion in
\cite{Danchev2020}). More specifically, within the new approach we
define the lower limit of integration as the lower value of the
radius $x$, corresponding to the left point of the half-width
$\Gamma$. To test the sensitivity of this criterion, we perform
additional calculations taking $\Gamma\pm 10\%$. In this case, the
results for the ratio $1/q$ of a given nucleus displayed a very
small sensitivity in the case of Bonn B and Bonn CD potentials,
while in the case of Brueckner and Skyrme EDFs the results when
applying criteria related to $\Gamma$ and $\Gamma\pm 10\%$ are
almost identical. We also note that in the new scheme there are no
singularities in the integrand of Eq.~(\ref{eq:25}) as those
mentioned for the integrand of Eq.~(\ref{eq:22}).

\begin{table*}
\caption{The ranges of changes of $1/q$ (scheme II) and $\kappa$
(scheme I) \cite{Antonov2016,Danchev2020} with Skyrme and
Brueckner EDFs and BHF method with Bonn B and Bonn CD potentials
for the Ni, Sn, and Pb isotopic chains.} \label{tab1}
\begin{center}
\begin{tabular}{ccccccccc}
\hline \hline \noalign{\smallskip}
&  Ni & & & Sn & & & Pb \\
\noalign{\smallskip}\hline\noalign{\smallskip}
& $1/q$ & $\kappa$ & & $1/q$ & $\kappa$ & & $1/q$ & $\kappa$ \\
\hline \noalign{\smallskip}
Skyrme    & 2.07-2.36 & 1.53-1.70 & & 1.63-2.37 & 1.58-2.02 & & 1.97-2.09 & 1.67-1.71 \\
Brueckner & 1.14-1.24 & 2.22-2.44 & & 0.94-1.16 & 2.40-2.90 & & 1.01-1.04 & 2.62-2.64 \\
Bonn B    & 1.03-1.08 & 1.80-1.90 & & 0.83-0.97 & 2.00-2.48 & & 0.84-0.88 & 2.54-2.80 \\
Bonn CD   & 1.01-1.06 & 1.80-2.00 & & 0.82-0.95 & 2.00-2.48 & & 0.81-0.83 & 2.54-2.80 \\
\noalign{\smallskip}\hline \hline
\end{tabular}
\end{center}
\end{table*}

\begin{figure*}
\centering
\includegraphics[width=0.8\linewidth]{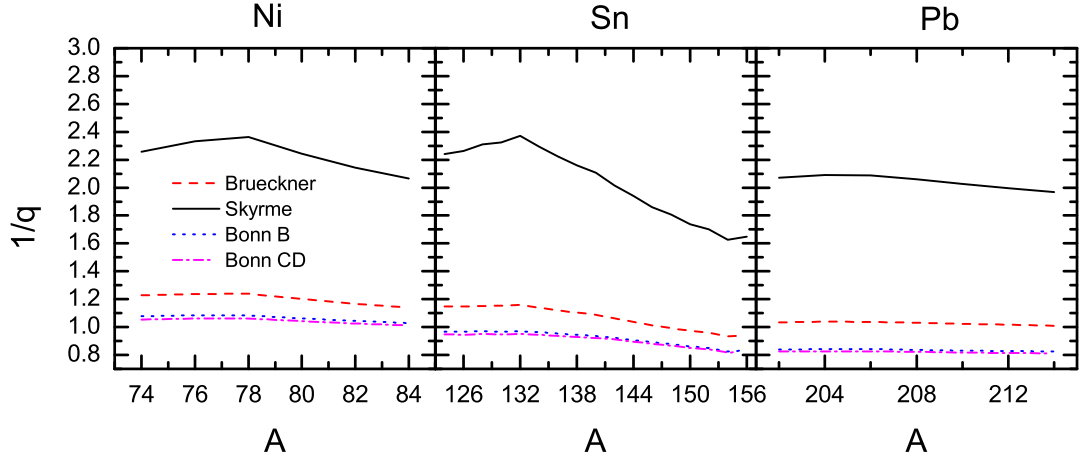}
\caption[]{(Color online) The quantity $1/q=S^{V}/S^{S}$
[following from Eq.~(\ref{eq:25})] as a function of $A$ for the
isotopic chains of Ni, Sn, and Pb obtained using Brueckner EDF
(dashed line), Skyrme EDF (solid line) and BHF method with Bonn B
(dotted line) and Bonn CD (dash-dotted line) potentials from
Refs.~\cite{Machleidt_ANP1989,Machleidt2001}. The weight function
$|F(x)|^{2}$ [Eq.~(\ref{eq:17})] used in the calculations is
obtained by means of the densities derived within a
self-consistent Skyrme-Hartree-Fock plus BCS method with SLy4
force. \label{fig3}}
\end{figure*}

Next, we show in Fig.~\ref{fig3} the results of the calculations
following from Eq.~(\ref{eq:25}) of the ratio $1/q=S^{V}/S^{S}$ as
a function of the mass number $A$ for the isotopic chains of Ni,
Sn, and Pb with SLy4 force. In Table~\ref{tab1} the values of this
ratio obtained within the new scheme are compared with the values
of $\kappa$ [Eq.~(\ref{eq:22})] calculated from our previous
scheme within the CDFM \cite{Antonov2016,Danchev2020}. We would
like to emphasize that this comparison is between quantities
obtained in two different CDFM schemes and it can serve basically
to show the influence and the importance of the preliminary
assumptions  and shortcomings made of scheme I and the advantage
of the new scheme that is free from them.

In general, the values of $1/q$ within the new CDFM scheme
calculated using the Skyrme EDF for the isotopic chains of Ni, Sn,
and Pb are between 1.70 and 2.40. This range of values is similar
to the estimations for $\kappa(A)$ [Eq.~(\ref{eq:22})] of
Danielewicz {\it et al.} obtained from a wide range of available
data on the binding energies \cite{Dan2003}, of Steiner {\it et
al.} \cite{Steiner2005}, and from a fit to other nuclear
properties, such as the excitation energies to IAS and skins
\cite{Dan2004}

$2.6 \leq \kappa \leq 3.0$

\noindent and from masses and skins \cite{Dan2004}

$2.0 \leq \kappa \leq 2.8$ .

\noindent The values of $1/q$ obtained using the Brueckner EDF for
the Ni isotopic chain with SLy4 force are in agreement partly with
that obtained in Ref.~\cite{Diep2007} by Dieperink and Van Isacker
from the analyses of masses and skins

$1.6 \leq \kappa \leq 2.0$ .

\noindent The obtained values of $1/q$ for Sn and Pb isotopes
using the Brueckner EDF together with the ones when using both
Bonn potentials are close to the value of 1.14 given by Bethe in
Ref.~\cite{Bethe71} and to the estimated value of 1.1838 by Myers
and Swiatecki \cite{Myers1969}. Generally, we can note that the
results of the new scheme for $1/q$, in particular using Skyrme
and Brueckner EDFs, cover reasonably the estimated values of
$\kappa$ (between 1.14 and 2.80) in a better way than in the
previous scheme.

Here we note the observed peaks in the ratio $1/q$ at $A=78$ and
$A=132$ for Ni and Sn isotopes, respectively. They are more
pronounced for the choice of the Skyrme EDF, less pronounced for
Brueckner EDF, and are somewhat smoothed out for Bonn B and Bonn
CD potentials. We attribute these peaks to the sharp nuclear
density transition when passing double-magic nuclei, such as
$^{78}$Ni and $^{132}$Sn, in an isotopic chain. The peculiarities
of $\rho(r)$ (and consequently the derivative of $\rho(r)$ which
determines the weight function $|F(x)|^{2}$) for the closed shells
lead to the existence of "kinks" that had been found and discussed
in our previous works
\cite{Gaidarov2011,Gaidarov2012,Antonov2016,Antonov2018,Danchev2020}.
In the case of Pb isotopic chain (see Fig.~\ref{fig3}) such kink
does not exist at $A=208$ and this reflects the smooth behavior
without kinks of $S(A)$ [Eq.~(\ref{eq:18})] and related quantities
for the Pb isotopic chain \cite{Gaidarov2011,Gaidarov2012}.
Similar peaks in the ratio $\kappa$ as a function of the mass
number have been observed in our previous studies
\cite{Antonov2016,Danchev2020}.

\begin{figure*}
\centering
\includegraphics[width=0.8\linewidth]{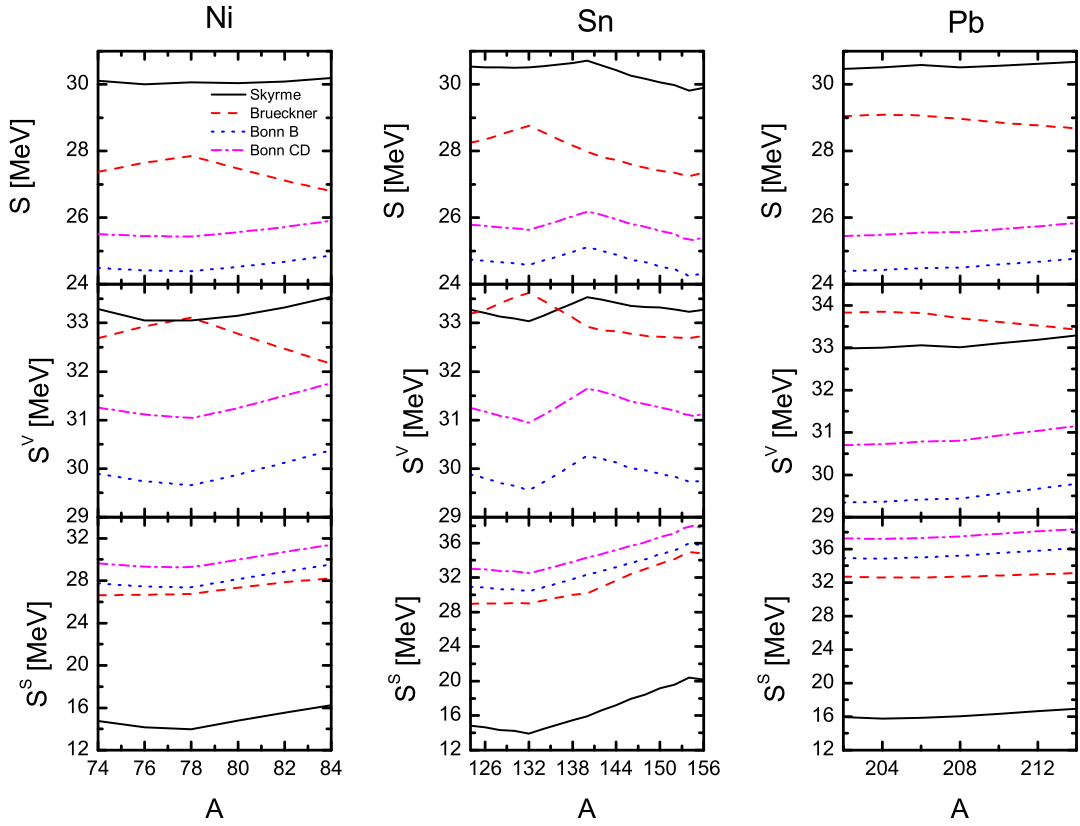}
\caption[]{(Color online) The symmetry energy $S$
[Eq.~(\ref{eq:18})] and its volume $S^{V}$ [Eq.~(\ref{eq:5})] and
surface $S^{S}$ [Eq.~(\ref{eq:6})] components for the isotopic
chains of Ni, Sn, and Pb obtained using Brueckner EDF (dashed
line), Skyrme EDF (solid line) and BHF method with Bonn B (dotted
line) and Bonn CD (dash-dotted line) potentials from
Refs.~\cite{Machleidt_ANP1989,Machleidt2001}. The weight function
$|F(x)|^{2}$ [Eq.~(\ref{eq:17})] used in the calculations is
obtained by means of the densities derived within a
self-consistent Skyrme-Hartree-Fock plus BCS method with SLy4
force. \label{fig4}}
\end{figure*}

The values of the symmetry energy $S$ [Eq.~(\ref{eq:18})] and its
volume $S^{V}$ [Eq.~(\ref{eq:5})] and surface $S^{S}$
[Eq.~(\ref{eq:6})] components as functions of $A$ deduced within
the new scheme for the same isotopic chains are presented in
Fig.~\ref{fig4}. The calculated symmetry energy for the three
isotopic chains and all considered potentials turns out to be
between 24 and 31 MeV (see Fig.~\ref{fig4}). In practice,
predictions for the symmetry energy vary substantially (28--38
MeV), e.g., an empirical value of the symmetry energy $30\pm4$ MeV
is given in Refs.~\cite{Han88,Nik2011}. The values of the volume
contribution $S^{V}$ to the NSE obtained within the new scheme in
the case of Brueckner and Skyrme EDFs are smaller than the ones
derived from the previous CDFM scheme I (presented in Tables~I and
III of Ref.~\cite{Antonov2016}). We would like to emphasize that
the results for $S^{V}$ in the scheme II (between 29 and 34 MeV)
are more realistic than the ones previously obtained within our
scheme I, for instance, using Brueckner EDF (between 41.5 and 43
MeV). The new results with scheme II are in good agreement with
the available phenomenological estimations, as follows:

Ref.~\cite{Dan2004}: $30.0 \leq S^{V} \leq 32.5$ MeV

Ref.~\cite{Danielewicz}: $31.5 \leq S^{V} \leq 33.5$ MeV .

\noindent In the case of Ni isotopic chain our previous
calculations \cite{Danchev2020} with SLy4 force provided values of
the volume symmetry energy within 27.6 and 28.1 MeV for Bonn B
potential and within 28.4--29.1 MeV for Bonn CD potential. In the
new approach for the same potentials the corresponding values of
$S^{V}$ are larger by 2 MeV and are better compared with the
results presented in Refs.~\cite{Dan2004,Danielewicz}. Concerning
the surface component of the NSE $S^{S}$, it is known that this
component is poorly constrained by empirical data. Therefore, it
is useful to test different EDFs and nuclear potentials within
different approaches to collect more information about it.
Figure~\ref{fig4} shows that the range of the values obtained for
$S^{S}$ and for Ni, Sn, and Pb isotopes in the case of Skyrme EDF
is 14--18 MeV. These results come closer to the limits on the
surface symmetry parameter 11 MeV$\leq \beta \leq$14 MeV
established in Ref.~\cite{Dan2003}. The new CDFM scheme gives
larger values for the surface component in the case of the three
other potentials (Brueckner, Bonn B, and Bonn CD).

We would like to note that the same peculiarities (as for the
ratio $1/q=S^{V}/S^{S}$ presented in Fig.~\ref{fig3}), namely
"kinks" appear in the cases of $S$, $S^{V}$, and $S^{S}$ as
functions of the mass number $A$ at the double-magic $^{78}$Ni and
$^{132}$Sn isotopes. They are stronger or weaker and depending on
the use of a given nuclear potential. In Fig.~\ref{fig4} a kink
appears for $S(A)$ and $S^{V}(A)$ not only for the double-magic
$^{132}$Sn but also for the semimagic $^{140}$Sn nucleus. As was
discussed in Ref.~\cite{Antonov2016}, the latter is related to the
closed $2f_{7/2}$ subshell for neutrons. Kinks of the $A$
dependence of the symmetry energy and its components in the Pb
isotopic chain are not observed.

To summarize this Section, we would like to stress that the
comparison of the results of the new scheme II with those of
scheme I is mainly informative to test the role of the
approximations in scheme I versus the new procedure in scheme II
that is free from them. This comparison together with the
comparison to phenomenological estimates of the NSE components,
mainly the volume one, allows us to conclude that the results of
scheme II are more realistic.

\section{Summary and conclusions}
\label{sec:conclusions}

The results of the present work can be summarized as follows:

i) We provide an alternative approach (scheme II) to calculate the
ratio $q(A)=S^{S}(A)/S^{V}(A)$ of the surface to volume components
of the NSE within the framework of the CDFM in a more direct and
simple way and having stronger physical grounds than the former
one (scheme I) that had been used in our previous works
\cite{Danchev2020,Antonov2016}. In the new approach we firstly
determine the ratio $q(x)$ for a flucton in the CDFM model from
the basic Droplet Model mass formula and then we use the
convolution of $q(x)$ with $|F(x)|^{2}$ to construct $q(A)$ for
finite nuclei following the standard CDFM procedure. In this way
the new scheme avoids some conceptual and mathematical
shortcomings that were met in the previous scheme.

ii) The results for $q(A)$ and the components $S^{S}(A)$ and
$S^{V}(A)$ are obtained from calculations based on Skyrme and
Brueckner energy-density functionals for nuclear matter and
nonrelativistic Brueckner-Hartree-Fock method with realistic Bonn
B and Bonn CD $NN$ potentials. As in our previous scheme, by
applying the CDFM the finite nuclei densities from the isotopic
chains of Ni, Sn, and Pb are obtained in self-consistent
Hartree-Fock+BCS calculations with SLy4 Skyrme effective
interaction.

iii) We would like to note the dependence of the results for the
ratio of $S^{S}$ to $S^{V}$ on the effective nuclear potentials
used in the calculations. In this respect, the results of our
calculations using Skyrme EDF turn out to be close to the
different estimations obtained from a fit to nuclear properties,
such as the excitation energies to IAS and neutron-skin thickness
\cite{Dan2004}, masses, and others. The values of $1/q$ obtained
using the Brueckner EDF for the Ni isotopic chain are in agreement
with those obtained in Ref.~\cite{Diep2007} from the analyses of
masses and skins. In the case of Bonn B and Bonn CD two-body
potentials the results for the ratio $1/q$ approach the estimated
values from the works of Bethe \cite{Bethe71} and Myers and
Swiatecki \cite{Myers1969}. Overall, the results of the new
scheme for $1/q$ cover reasonably the whole region of estimated
values for $\kappa$ (between 1.14 and 2.80) and in some cases are
somewhat better than the values obtained in the previous scheme.

iv) The values of the symmetry energy $S$ for the three isotopic
chains and all considered potentials are between 24 and 31 MeV
that is in accordance with the region of its empirical values $30
\pm 4$ MeV given in Refs.~\cite{Han88,Nik2011}. The results for
the volume component $S^{V}(A)$ of NSE in scheme II (between 29
and 34 MeV) are in good agreement with those of
Refs.~\cite{Dan2004,Danielewicz} (between 30 and 33.5 MeV). The
values of the surface contribution $S^{S}(A)$ in scheme II in
the case of Skyrme EDF (14-18 MeV) come closer to the region of
11-14 MeV established in Ref.~\cite{Dan2003}.

v) Analyzing the isotopic sensitivity of $S^{V}(A)$, $S^{S}(A)$,
and their ratio $1/q(A)$ we observe peculiarities ("kinks") of
these quantities as functions of the mass number $A$ in the cases
of the double-magic $^{78}$Ni and $^{132}$Sn isotopes, as well as
a "kink" of $S^{V}(A)$ for $^{140}$Sn. No pronounced peak at the
double-magic nucleus with $A=208$ in the Pb chain is found. The
mentioned peculiarities in the behavior of the corresponding
curves for the same quantities have been observed also in our
previous CDFM scheme.

Finally, we point out that the results for the NSE components and
their ratio obtained within the two CDFM schemes are comparable in
many considered cases and cover a range of values that is
compatible with the range of available empirical data and with
other theoretical results, showing the power of the CDFM method,
that includes effects of nucleon-nucleon correlations of
collective type. However, we would like to emphasize that the
presented comparison of the results of both schemes is informative
mainly for the role of the approximations made in scheme I, while
scheme II is free from those approximations and is considered to
be more reliable and realistic leading to results that are in
better agreement with data.

\begin{acknowledgments}
A.N.A, M.K.G., and D.N.K. are grateful for the support of the
Bulgarian National Science Fund under Contract No.~KP-06-N38/1.
I.C.D also wishes to acknowledge partial financial support from
the University of Mount Olive Professional Development Fund. P.S.
acknowledges support from Ministerio de Ciencia e Innovaci\'on
MCIU/AEI/FEDER,UE (Spain) under Contract No. PGC2018-093636-B-I00.
\end{acknowledgments}

\end{document}